\documentclass[aps,pre,twocolumn,10pt,superscriptaddress,showpacs]{revtex4-1}

\def\bra#1{\mathinner{\langle{#1}|}}
\def\ket#1{\mathinner{|{#1}\rangle}}
\def\braket#1{\mathinner{\langle{#1}\rangle}}

\let\protect\relax
{\catcode`\|=\active
  \xdef\Braket{\protect\expandafter\noexpand\csname Braket \endcsname}
  \expandafter\gdef\csname Braket \endcsname#1{\begingroup
     \ifx\SavedDoubleVert\relax
       \let\SavedDoubleVert\|\let\|\BraDoubleVert
     \fi
     \mathcode`\|32768\let|\BraVert
     \left\langle{#1}\right\rangle\endgroup}
}
\def\BraVert{\@ifnextchar|{\|\@gobble}
     {\egroup\,\mid@vertical\,\bgroup}}
\def\BraDoubleVert{\egroup\,\mid@dblvertical\,\bgroup}
\let\SavedDoubleVert\relax

{\catcode`\|=\active
  \xdef\set{\protect\expandafter\noexpand\csname set \endcsname}
  \expandafter\gdef\csname set \endcsname#1{\mathinner
        {\lbrace\,{\mathcode`\|32768\let|\midvert #1}\,\rbrace}}
  \xdef\Set{\protect\expandafter\noexpand\csname Set \endcsname}
  \expandafter\gdef\csname Set \endcsname#1{\left\{%
     \ifx\SavedDoubleVert\relax \let\SavedDoubleVert\|\fi
     \:{\let\|\SetDoubleVert
     \mathcode`\|32768\let|\SetVert
     #1}\:\right\}}
}
\def\midvert{\egroup\mid\bgroup}
\def\SetVert{\@ifnextchar|{\|\@gobble}
    {\egroup\;\mid@vertical\;\bgroup}}
\def\SetDoubleVert{\egroup\;\mid@dblvertical\;\bgroup}

\begingroup
 \edef\@tempa{\meaning\middle}
 \edef\@tempb{\string\middle}
\expandafter \endgroup \ifx\@tempa\@tempb
 \def\mid@vertical{\middle|}
 \def\mid@dblvertical{\middle\SavedDoubleVert}
\else
 \def\mid@vertical{\mskip1mu\vrule\mskip1mu}
 \def\mid@dblvertical{\mskip1mu\vrule\mskip2.5mu\vrule\mskip1mu}
\fi

\usepackage{graphicx}
\usepackage[T1]{fontenc}
\usepackage[]{amsfonts}
\usepackage[]{amsmath}
\usepackage[]{rotating}
\usepackage[]{color}
\usepackage[]{float}
\usepackage[]{fancyhdr}
\usepackage[]{booktabs}
\usepackage[]{epsfig}
\usepackage{appendix}
\usepackage{amssymb}
\usepackage{psfrag}
\usepackage{setspace}

\begin{document}

\title{Quantum driving and work}

\author{J. Salmilehto}
\affiliation{QCD Labs, COMP Centre of Excellence, Department of Applied Physics, Aalto University, P.O. Box 13500, FI-00076 Aalto, Finland}
\author{P. Solinas}
\affiliation{SPIN-CNR, Via Dodecaneso 33, I-16146 Genova, Italy}
\author{M. M\"ott\"onen}
\affiliation{QCD Labs, COMP Centre of Excellence, Department of Applied Physics, Aalto University, P.O. Box 13500, FI-00076 Aalto, Finland}
\affiliation{Low Temperature Laboratory (OVLL), Aalto University, P.O. Box 13500, FI-00076 Aalto, Finland}


\begin{abstract}

As quantum systems become more experimentally accessible, we are forced to reconsider the notions of control and work to fully account for quantum effects. To this end, we identify the work injected into a quantum system during a general quantum-mechanical driving protocol and quantify the relevant heat flows. The known results that are applicable in the limit of a classical drive are shown to emerge from our equations as a special case. Using the established framework, we show that the Bochkov--Kuzovlev identity for the exclusive work distribution is modified in a nontrivial way by the accumulation of system--drive correlations resulting from quantum backaction. Our results accentuate the conceptual and discernible differences between a fully quantum-mechanical and classical driving protocols of quantum systems.

\end{abstract}

\maketitle

\section{Introduction}

The past decade has witnessed great developments in the measurement precision and degree of control of nanoscale physical systems. Such major milestones have been achieved, for example, in the frameworks of quantum information processing~\cite{prl100/247001, prl109/060501, nature489/541, nature496/334} and fluctuation relations in steered evolution~\cite{rmp83/771, prl109/180601, np9/644}. This progress calls for a detailed and highly accurate theory for driven systems.

In the typical approach to driving a quantum system, the closed-system dynamics are influenced by a time-dependent Hamiltonian stemming from the action of an external driving force~\cite{rmp81/1665, rmp83/771}. The action defines the driving protocol and, hence, quantifies temporal changes related to it. Despite its usefulness, the approach entails the underlying assumption that the time-dependent control is carried out by an external classical entity and, consequently, it is potentially omitting important physical phenomena governed by the ubiquitous quantum-mechanical backaction. Including driving as a fully quantum process would not only provide fundamental understanding of the physics but would also open a new path in treating and designing driven quantum systems.

Classical thermodynamics~\cite{thermo, clasmech} motivates a long-standing question related to driving: how does one identify the work performed on a quantum system during a driving protocol and, subsequently, calculate possible energy flow to a coupled heat bath? Even within the typical approach to driving, answering this question has proven a formidable task as work relates to a process rather than a time-local quantum observable~\cite{pre88/032146, prl102/210401, jpsj69/2367, prl90/170604, prl93/048302, pre71/066102, pre73/046129, epl79/10003, pre75/050102, jsm2009/P02025, prl100/230404, prb87/060508}. This problem is made even more interesting by the close connection it shares with nonequilibrium work relations where sampling over the stochastic ensemble generated by realizations of the dynamics provides information about the equilibrium properties of the system~\cite{rmp81/1665, rmp83/771, prl102/210401, pre88/032146}. High-precision measurements of such relations would necessarily have to account for the system--drive backaction.

In this paper, we treat the driving on an equal footing with the rest of the quantum dynamics by dividing the total system into its constituent parts and assigning the driving protocol to a specific subsystem. This composite framework facilitates quantum backaction between the system and the drive allowing us to study driving in much greater detail than with the assumption of external classical drive. We propose a natural definition for the injected work and show how the intracomposite energy transfer including heat flows can be evaluated. Taking the total system to the classical driving limit re-establishes the previously employed driving framework and we retrieve the known result for the injected work~\cite{prb87/060508}. In addition, we employ a stochastic approach and show that using the full composite picture modifies the Bochkov--Kuzovlev identity~\cite{spjetp45/125, rmp83/771} due to the accumulation of system--drive correlations resulting from the backaction. In the classical driving limit, such correlations vanish and we recover the usual Bochkov-Kuzovlev identity.

\section{Quantum driving and work injection}

\begin{figure}
\centering
\includegraphics[width=7cm]{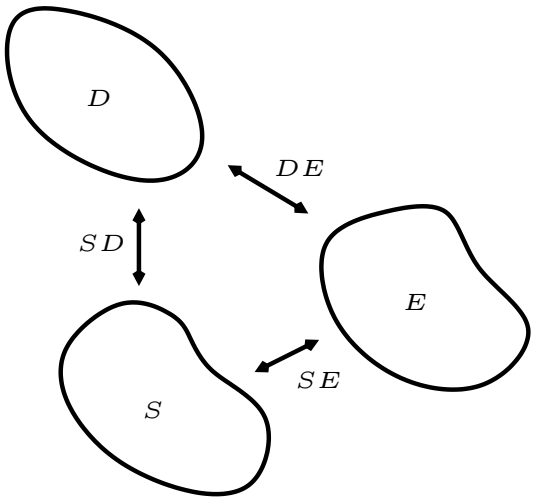}
\caption{Schematic representation of the total composite system implementing quantum driving. $S$ denotes the subsystem of interest, $D$ denotes the drive subsystem, and $E$ denotes the environment. Furthermore, $SD$, $SE$ and $DE$ denote the interactions between the subsystems.}
\label{fig:Figsupp1}
\end{figure}

To specify the process of driving, we divide the total quantum system into its constituent interacting parts: the subsystem of interest $S$, the drive $D$, and the environment $E$. The total Hamiltonian is expressed as
\begin{equation}
\begin{split}
\hat{H} =& \hat{I}_D \otimes \hat{H}_S \otimes \hat{I}_E + \hat{H}_D \otimes \hat{I}_S \otimes \hat{I}_E + \hat{I}_D \otimes \hat{I}_S \\ &\otimes \hat{H}_E + \hat{H}_{SD} \otimes \hat{I}_E + \hat{I}_D \otimes \hat{H}_{SE} + \hat{H}_{DE},
\end{split}
\label{eq:H}
\end{equation}
where $\hat{I}_i$ is the identity operator in the Hilbert space of the $i$th subsystem, $\hat{H}_i$ is the corresponding Hamiltonian and $\hat{H}_{ij}$ is the interaction Hamiltonian between the $i$th and the $j$th subsystems. Figure~\ref{fig:Figsupp1} provides a schematic representation of the subsystem division of the composite.
We do not employ external, time-dependent parameters on any of the Hamiltonians but assume that the full information of the driving protocol is encoded in the internal interactions of the composite and the initial state of $D$. To this end, we define the driving done to $S$ as the dynamics induced by the interaction with the desired drive degrees of freedom. The environment accounts for all interactions with the system and the drive that do not facilitate such ideal reduced system dynamics. 

Within this line of thinking, we propose a natural definition for the average work $W_Q$ injected into the system during the driving protocol: the change in the internal energy of the system and its interaction energy with the drive combined with the energy dissipated directly from the system to the environment. This approach extends the notion of the dynamical work agent~\cite{pre86/011111} related to the work source in classical Hamiltonian dynamics~\cite{prx3/041003} to a fully consistent quantum description. Importantly, the drive does not have to be rapidly self-equilibrating and quantum backaction is allowed. The instantaneous injected power corresponding to the definition is explicitly derived in Appendix~\ref{sec:ader} and takes the form
\begin{equation}
\begin{split}
\frac{d}{dt} W_Q = -\frac{i}{\hbar} \mathrm{Tr}_{S+D} \{ \hat{\rho}_{SD}[\hat{H}_{SD},\hat{H}_D \otimes \hat{I}_S] \},
\end{split}
\label{eq:WQ}
\end{equation}
where $\hat{\rho}_{SD} = \mathrm{Tr}_E \{ \hat{\rho} \}$, $\hat{\rho}$ is the total density operator, $\mathrm{Tr}_E$ is the trace over the environment, and $\mathrm{Tr}_{S+D}$ is the combined trace over the system and the drive. Importantly, Eq.~(\ref{eq:WQ}) has a trace over $S$ and $D$ instead of the total composite space and the information on the heat dissipation is confined within the reduced density operator $\hat{\rho}_{SD}$. In addition, we can write the average power dissipated directly from the system as [see Appendix~\ref{sec:ader}]
\begin{equation}
\begin{split}
\frac{d}{dt} Q_S = &-\frac{i}{\hbar} \mathrm{Tr}_{S+E} \{ \hat{\rho}_{SE} [\hat{H}_{SE},\hat{H}_S \otimes \hat{I}_E] \} \\ &-\frac{i}{\hbar} \mathrm{Tr} \{ \hat{\rho} [\hat{I}_D \otimes \hat{H}_{SE} + \hat{H}_{DE}, \hat{H}_{SD} \otimes \hat{I}_E] \},
\end{split}
\label{eq:dQ/dt}
\end{equation}
where $Q_S$ is the corresponding dissipated heat and $\mathrm{Tr}$ is the trace over the total composite Hilbert space. In general, there can also be energy transfer caused by the direct interaction between $D$ and $E$, $Q_D$, that must be accounted for to calculate the total dissipated heat $Q_{\mathrm{tot}} = Q_S + Q_D$. The corresponding heating power assumes the form [see Appendix~\ref{sec:ader}]
\begin{equation}
\begin{split}
\frac{d}{dt} Q_D = -\frac{i}{\hbar} \mathrm{Tr} \{ \hat{\rho} [\hat{H}_{DE}, \hat{H}_D \otimes \hat{I}_S \otimes \hat{I}_E] \}.
\end{split}
\label{eq:dQ_D/dt}
\end{equation}
Conservation of energy dictates that $-\Delta \braket{\hat{H}_D} = W_Q + Q_D$ implying that the injected work describes the beneficial energy transfer in the driving process and is not necessarily equivalent to the total energy extracted from the drive. Note that $W_Q$ defined in Eq.~(\ref{eq:WQ}) includes the change in the system--drive interaction energy and, hence, corresponds to the average {\it inclusive work} in the typical notation~\cite{rmp81/1665, rmp83/771}.

For practical applications, the reduced system dynamics are typically accessed in the form of a general master equation $\dot{\hat{\rho}}_{SD} = -(i/ \hbar) [\hat{H}_{SD}',\hat{\rho}_{SD}] + \hat{\mathcal{D}}$, where $\hat{H}_{SD}' = \hat{I}_D \otimes \hat{H}_S + \hat{H}_D \otimes \hat{I}_S + \hat{H}_{SD}$ and the dissipator $\hat{\mathcal{D}}$ accounts for the effect of the environment on the system-drive dynamics~\cite{tToOQS, QDS}. Using this representation and applying the general definition of operator current~\cite{pra85/032110} to Eqs.~(\ref{eq:dQ/dt}) and (\ref{eq:dQ_D/dt}), the dissipated-heat terms are recast into
\begin{equation}
\begin{split}
\frac{d}{dt} Q_S = - \mathrm{Tr}_{S+D} \{ \hat{\mathcal{D}} (\hat{I}_D \otimes \hat{H}_S + \hat{H}_{SD}) \},
\label{eq:masterheating}
\end{split}
\end{equation}
and
\begin{equation}
\begin{split}
\frac{d}{dt} Q_D = - \mathrm{Tr}_{S+D} \{ \hat{\mathcal{D}} (\hat{H}_D \otimes \hat{I}_S) \}.
\end{split}
\label{eq:masterdriveheating}
\end{equation}
These results along with Eq.~(\ref{eq:WQ}) provide a means for exploring physical systems whose description likely requires the use of approximative dissipative methods.

\section{Classical driving limit}

The concept of \textit{classical driving} of a quantum system emerges from the full composite picture given above when we assume that the drive acts as a classical entity, that is, its state is unaffected by the internal interactions of the composite. As a result, it and the rest of the composite remain uncorrelated at all times so that the Born approximation~\cite{tToOQS} is valid $\hat{\rho} = \hat{\rho}_D \otimes \hat{\rho}_{SE}$. In addition, we omit the drive--environment interaction described by $\hat{H}_{DE}$ to retrieve the typical classical picture. Within these assumptions, the system-environment dynamics are determined by [see Appendix~\ref{sec:aclas} for a full derivation]
\begin{equation}
\begin{split}
\hat{H}_{CL} = \hat{H}_{CL,S} \otimes \hat{I}_E + \hat{I}_S \otimes \hat{H}_E + \hat{H}_{SE},
\end{split}
\label{eq:HCL}
\end{equation}
where $\hat{H}_{CL,S} = \hat{H}_S + \sum_{\alpha} \hat{A}_{\alpha} \braket{\hat{B}_{\alpha}}_D$. Here the expectation value is defined by $\braket{\hat{B}_{\alpha}}_D = \mathrm{Tr}_D \{ \hat{\rho}_D \hat{B}_{\alpha} \}$ and we exploit the general decomposition of the interaction Hamiltonian $\hat{H}_{SD} = \sum_{\alpha} \hat{B}_{\alpha} \otimes \hat{A}_{\alpha}$, where $\hat{B}_{\alpha} = \hat{B}_{\alpha}^{\dagger}$ and $\hat{A}_{\alpha} = \hat{A}_{\alpha}^{\dagger}$~\cite{tToOQS}. The usual time-dependence of the system Hamiltonian in driving is introduced by $\braket{\hat{B}_{\alpha}}_D$ which is determined by the time-evolution of the uncorrelated drive subsystem. Our definition of a classical drive corresponds to that used in Refs.~[\onlinecite{epjst151/181},~\onlinecite{epl83/30008}] for a bipartite system in the limit of vanishing mutual corrections to the drive dynamics. We define the classical drive in this limit to assert it as an external independent work source as detailed in Appendix~\ref{sec:aclas}. Furthermore, the cross-over from exact reduced dynamics to a parametric field dependence of a qubit is studied for the spin-$\frac{1}{2}$ star with frustration in Ref.~\cite{pnas110/6748} using the so-called generalized coherent state formalism~\cite{rmp62/867, epl79/40003}.

In analogy with the fully classical case~\cite{prl78/2690}, the classically injected power can be written with the help of a power operator $\hat{P}_S = (\partial \hat{H}'_{CL,S} [\lambda (t)] / \partial \lambda) (\partial \lambda / \partial t)$~\cite{prb87/060508}, where $\hat{H}'_{CL,S} [\lambda (t)]$ denotes the classically driven system Hamiltonian and $\lambda(t)$ is the time-dependent control parameter describing the effect of an external macroscopic work source on the system~\cite{pre71/066102, pre73/046129, epl79/10003, pre75/050102, jsm2009/P02025, rmp81/1665, prl102/210401, rmp83/771}. The resulting average injected power is $dW_{CL}/dt = \mathrm{Tr}_S \{ \hat{\rho}_S \hat{P}_S \}$ with $W_{CL}$ being the average injected work~\cite{prb87/060508}. For the classical driving assumption described above, Eq.~(\ref{eq:WQ}) reduces to $dW_{CL}/dt$ when we identify $\hat{H}'_{CL,S} [\lambda (t)] = \hat{H}_{CL,S}$ [see Appendix~\ref{sec:aclas} for details] yielding the known result for classically injected power~\cite{prb87/060508}
\begin{equation}
\begin{split}
\frac{d}{dt} W_{CL} &= \frac{d}{dt} \braket{\hat{H}_{CL,S}} + \frac{i}{\hbar} \mathrm{Tr}_{S+E} \{ \hat{\rho}_{SE}[\hat{H}_{CL,S} \otimes \hat{I}_E,\hat{H}_{SE}] \} \\ &= \frac{d}{dt} \braket{\hat{H}_{CL,S}} - \mathrm{Tr}_S \{ \hat{\mathcal{D}} \hat{H}_{CL,S} \},
\label{eq:classicalpower}
\end{split}
\end{equation}
where the injected power is immediately separated into the change in the internal energy of the effective system given by the first term and the system related heating power given by the second term after each equality.

\section{Example: Jaynes--Cummings model}

To analyse the quantum work injection with and without the classical driving assumption, we study as an example the resonant single-mode Jaynes--Cummings model~\cite{jmo40/1195} where the driving of the two-level system (TLS) is induced by the photonic bath. The system-drive Hamiltonian is
\begin{equation}
\begin{split}
\hat{H}^{JC} = \frac{\hbar\omega}{2} \hat{I}_D \otimes \hat{\sigma}_z + \hbar \omega \hat{b}^{\dagger} \hat{b} \otimes \hat{I}_S + \hbar g (\hat{b} \otimes \hat{\sigma}_+ + \hat{b}^{\dagger} \otimes \hat{\sigma}_-),
\end{split}
\label{eq:HJC}
\end{equation}
where $\hat{\sigma}_z = \ket{e}\bra{e} - \ket{g}\bra{g}$, $\hat{\sigma}_+ = \hat{\sigma}_-^{\dagger} = \ket{e}\bra{g}$, $\hat{b}^{\dagger}$ ($\hat{b}$) is the photonic creation (annihilation) operator, $\omega$ is the resonance angular frequency and $g$ is the TLS--drive coupling strength. The states $\ket{g}$ and $\ket{e}$ are the ground and excited states of the TLS, respectively. Using Eq.~(\ref{eq:HCL}), the corresponding classical driving Hamiltonian is given by
\begin{equation}
\begin{split}
\hat{H}_{CL,S}^{JC} = \frac{\hbar\omega}{2} \hat{\sigma}_z + e^{-i\omega t}S\hat{\sigma}_+ + e^{i\omega t}S^*\hat{\sigma}_-,
\end{split}
\label{eq:HJCCL}
\end{equation}
where $S = \hbar g \sum_{n=0}^{\infty} a_n^*a_{n+1}\sqrt{n+1}$ for the initial photonic state $\sum_{n=0}^{\infty} a_n \ket{n_F}$, where $\ket{n_F}$ is the $n$th Fock state. The time-dependent exponential terms result from the uncorrelated time-evolution of the drive. 

Comparison of Eqs.~(\ref{eq:HJC}) and (\ref{eq:HJCCL}) readily indicates that the classical driving assumption potentially ignores relevant dynamics depending on the specifics of the initial photonic state. For example, if we assume that the $S-D$ composite evolves unitarily starting from $\ket{n_F} \otimes \ket{e}$, the injected work in the corresponding classical driving $W_{CL,FOCK}$ vanishes along with the effective interaction. In quantum driving however, it can be solved using Eq.~(\ref{eq:WQ}) and the exact solution of the Jaynes-Cummings model~\cite{QO} as $W_{Q,FOCK}(t) = -\hbar \omega \sin^2 (\Omega_n t)$, where the Rabi frequency is $\Omega_n = g \sqrt{n+1}$. If the initial state is $\ket{\alpha} \otimes \ket{e}$, where $\ket{\alpha} = e^{-|\alpha|^2/2} \sum_{n=0}^{\infty} \frac{\alpha^n}{(n!)^{1/2}}\ket{n_F}$ is a coherent state with $\bar{n} = |\alpha|^2$ as the average photon number, the classical driving results in $W_{CL,COH} (t) = -\hbar \omega \sin^2 (g|\alpha|t)$. The respective work in quantum driving is $W_{Q,COH}(t) = -\hbar \omega e^{-|\alpha |^2} \sum_{n=0}^{\infty} \frac{|\alpha |^{2n}}{n!} \sin^2 (\Omega_n t)$ exhibiting the expected behavior of collapses and revivals of the Rabi oscillations~\cite{QO}. Note that the classically injected work approximates the one given by the composite approach well if we study times below the characteristic time $t_q = |\alpha|/g$ and the average photon number is much greater than unity.

\begin{figure}
\centering
\includegraphics[width=8.6cm]{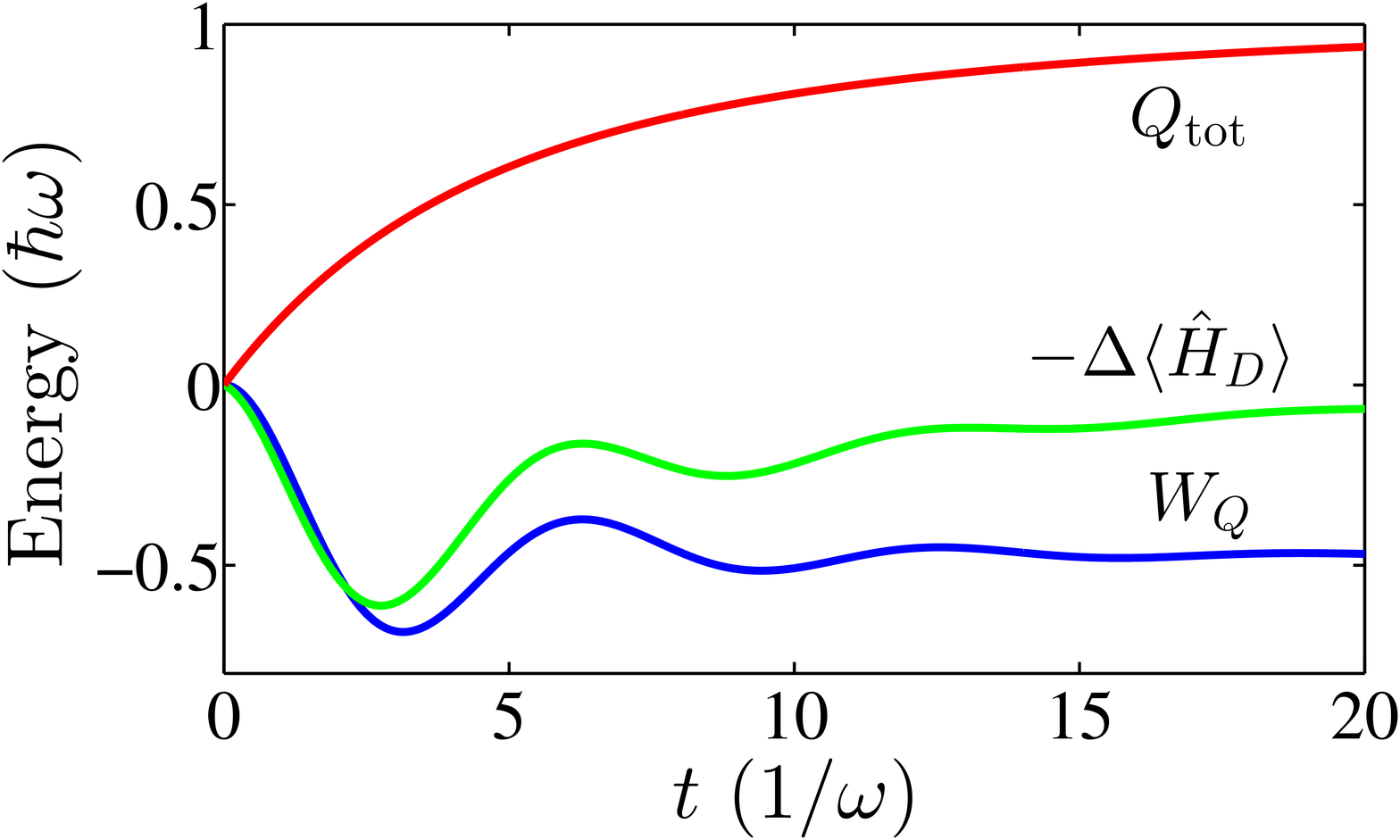}
\caption{(color online) Transferred energies in the Jaynes--Cummings model when the initial state is a product state of the excited two-level state and a Fock state with $n=0$ as a function of time. Blue curve shows the injected work $W_Q$, green curve shows the energy extracted from the drive $W_Q + Q_D = -\Delta \braket{\hat{H}_D}$, and red curve shows the total dissipated heat $Q_{\mathrm{tot}}$. We use $g=0.5 \omega$ for the coupling strength in Eq.~(\ref{eq:HJC}) and $\Theta = 0.2 \omega$ for the transition rate coefficient defined in the main text.}
\label{fig:Fig1}
\end{figure}

To study the effect of including a dissipative environment, we assume in our example that the system-drive composite is coupled to a reservoir inducing Markovian decay such that the dissipator is given by $\hat{\mathcal{D}} = \sum_{\substack{\text{$\varphi,\phi$}\\\text{$\epsilon_{\phi} > \epsilon_{\varphi}$}}} \{ 2\hat{L}_{\varphi\phi} \hat{\rho}_{SD} \hat{L}_{\varphi\phi}^{\dagger} - \hat{L}_{\varphi\phi}^{\dagger} \hat{L}_{\varphi\phi} \hat{\rho}_{SD} - \hat{\rho}_{SD} \hat{L}_{\varphi\phi}^{\dagger} \hat{L}_{\varphi\phi} \}$~\cite{tToOQS}. The Lindblad operators are defined as $\hat{L}_{\varphi\phi} = \sqrt{\gamma_R(\varphi,\phi)} \ket{\varphi_{JC}}\bra{\phi_{JC}}$, where $\hat{H}^{JC}\ket{\varphi_{JC}/\phi_{JC}} = \epsilon_{\varphi/\phi}\ket{\varphi_{JC}/\phi_{JC}}$. We assume that the dissipative dynamics are dominated by the system--environment interaction and take $\hat{H}_{SE} = \Gamma (\ket{g}\bra{e} + \ket{e}\bra{g}) \otimes \hat{E}$ where $\hat{E}$ is the environment part of the coupling operator. The transition rates are evaluated utilizing Fermi's golden rule so that $\gamma_R (\varphi,\phi) = \Theta |\braket{\varphi_{JC}|(\hat{I}_D \otimes \ket{g}\bra{e})|\phi_{JC}} + \braket{\varphi_{JC}|(\hat{I}_D \otimes \ket{e}\bra{g})|\phi_{JC}}|^2$, where $\Theta = |\Gamma|^2 S_E/\hbar^2$, and we assume white noise spectrum $S_E$ for simplicity. This evaluation in association with the Lindblad form enables direct heat transfer from the drive. We show the relevant energy changes in Fig.~\ref{fig:Fig1} obtained by placing an excitation in the TLS and assuming that the photonic system is initially in the vacuum state.
The decay rates are taken to be sufficiently large so that the usual oscillatory behavior caused by the transfer of the excitation is swiftly damped by the coupling to the bath and $Q_{\mathrm{tot}}$ approaches unity as the environment absorbs the excitation. Similarly $\Delta \braket{\hat{H}_D}$ approaches zero as the initial energy received from the TLS is dissipated. The injected work assumes a negative value describing energy transfer from the system to the drive and its asymptotic value is nonzero due to the direct heat transfer from the drive.

\section{Modification of the Bochkov-Kuzovlev identity}

Let us study the non-equilibrium work relations in our framework and employ the well-established two-measurement approach (TMA)~\cite{arxiv/0009244, arxiv/0007360, epl79/10003, jsm2009/P02025, rmp81/1665, rmp83/771, pre88/032146} for a general non-degenerate $N$-level system. For simplicity, we assume here that the system-drive composite is decoupled from the environment during the driving protocol. An extension of the TMA to strong system--bath couplings exists leading to partition functions for open systems~\cite{prl102/210401, rmp83/771}, but the extension still exploits the strictly classical driving as described earlier. As a prerequisite, we assume that we can probe the composite only by projective measurements of $\hat{H}_S$. This implies that the TMA only grants us access to the {\it exclusive work} $W_{\mathrm{excl}}$ not accounting for the change in the system--drive interaction energy~\cite{rmp83/771}. On the other hand, applying the TMA to joint system--drive measurements of $\hat{H}_S + \hat{H}_{SD}$ would result in trivial stochastics and no thermodynamic information would be retrieved. The typical quantum fluctuation relations avoid this problem by assuming classical driving~\cite{rmp81/1665, rmp83/771} which deposits the information of the driving protocol in the system degrees of freedom and, hence, allows the thermodynamic quantities to be acquired by measurements of the system alone.

We assume that the state of the composite system prior to the first measurement is $\hat{\rho}_{SD} (0) = \hat{\rho}_D (0) \otimes \hat{\rho}_S (0)$ and measure $\hat{H}_S$ at time $t=0$ so that the composite state collapses to $\hat{\rho}_{SD}^M (0) = \hat{\rho}_D (0) \otimes \ket{n^S}\bra{n^S}$ with probability $\rho_{S,nn} (0) = \braket{n^S|\hat{\rho}_S(0)|n^S}$ where $\epsilon_n\ket{n^S} = \hat{H}_S \ket{n^S}$. The total temporal evolution is given by $\hat{\rho}_{SD} (t) = \hat{U}(t,0) \hat{\rho}_{SD}^M (0) \hat{U}^{\dagger}(t,0)$ where $\hat{U}(t,0)$ is the unitary time-evolution operator for the system-drive composite. We also denote the reduced system density operator by $\hat{\rho}_S (t) = \mathrm{Tr}_D \{\hat{\rho}_{SD} (t) \}$. The second measurement at $t=T$ yields $\epsilon_k$ with probability
\begin{equation}
\begin{split}
P_{k,n}  = \mathrm{Tr}_D \{ \braket{k^S|\hat{U}(T,0) \hat{\rho}_D (0) \otimes \ket{n^S}\bra{n^S} \hat{U}^{\dagger}(T,0)| k^S} \}.
\end{split}
\label{eq:Pkn}
\end{equation}
The averaged exponentiated exclusive work is then written summing over the discrete stochastic trajectories as $\braket{e^{-\beta W_{\mathrm{excl}}}} = \sum_n \rho_{S,nn}(0) \sum_k P_{k,n} e^{-\beta (\epsilon_k-\epsilon_n)}$ where $\beta$ is the inverse temperature and $\braket{\dots}$ denotes the ensemble average over the distribution of exclusive work.

We assume that the system is initially in a Gibbs state $\hat{\rho}_S (0) = e^{-\beta \hat{H}_S}/Z_S$, where $Z_S = \mathrm{Tr}_S \{ e^{-\beta \hat{H}_S} \}$ is the partition function of the bare system~\cite{rmp81/1665, rmp83/771, rpp75/126001}. With this assumption, we obtain
\begin{equation}
\begin{split}
\braket{e^{-\beta W_{\mathrm{excl}}}} =& \sum_k \frac{e^{-\beta \epsilon_k}}{Z_S} \sum_n P_{k,n} \\ =& \sum_k \frac{e^{-\beta \epsilon_k}}{Z_S} \mathrm{Tr}_D \{ \langle k^S| \hat{U}(T,0) \hat{\rho}_D (0) \\ &\otimes \hat{I}_S \hat{U}^{\dagger}(T,0)|k^S \rangle \},
\end{split}
\label{eq:eW1}
\end{equation}
where we applied $\sum_n \ket{n^S}\bra{n^S} = \hat{I}_S$ after the second equality. Note that $\sum_k P_{k,n} = \mathrm{Tr}_{S+D} \{ \hat{\rho}_{SD} (T) \} = 1$. Furthermore, the classical driving assumption results in a factorizable time-evolution operator in Eq.~(\ref{eq:Pkn}) and, hence, $\sum_n P_{k,n} = 1$. By Eq.~(\ref{eq:eW1}), this assumption implies $\braket{e^{-\beta W_{\mathrm{excl}}}} = 1$ retrieving the usual Bochkov-Kuzovlev identity~\cite{spjetp45/125, rmp83/771}. Beyond the classical driving assumption, $\braket{e^{-\beta W_{\mathrm{excl}}}}$ is generally not unity but its value depends on the system--drive correlations accumulated during the driving protocol. We can additionally write $\braket{e^{-\beta W_{\mathrm{excl}}}} = Z_S'(T)/Z_S'(0)$ if we define $Z_S'(t) = \mathrm{Tr}_S \{ \mathrm{Tr}_D \{ e^{-\beta \hat{H}_S^H(t)} \hat{\rho}_D(0) \} \}$ where $\hat{H}_S^H(t) = \hat{U}^{\dagger}(t,0) \hat{I}_D \otimes \hat{H}_S\hat{U}(t,0)$ is the system Hamiltonian in the complete Heisenberg picture. Thus, Eq.~(\ref{eq:eW1}) can be interpreted in terms of a partition function corresponding to the temporal evolution of the drive-averaged canonical state of the system. Details and discussion on this interpretation are presented in Appendix~\ref{sec:aint}. The modification of the Bochkov-Kuzovlev identity serves to highlight the limited range of validity of the usual quantum fluctuation relations. It should be observable in physical systems where the backaction is not negligible and the system energy can be measured with sufficient accuracy. Physical prerequisities for observing the modification are potentially emergent in fields such as circuit~\cite{nature431/162} and cavity~\cite{rmp73/565} quantum electrodynamics, solid state qubits~\cite{nature398/786, nature453/1031}, and trapped ions~\cite{rmp75/281}.

\subsection{Jaynes--Cummings model revisited}

Let us return to the Jaynes--Cummings model presented above and take the initial state of $D$ to be $\sum_{n=0}^{\infty}a_n \ket{n_F}$. Thus, Eq.~(\ref{eq:eW1}) takes the form $\braket{e^{-\beta W_{\mathrm{excl}}}} = 1 + \frac{1}{Z_S} (e^{-\beta \hbar \omega /2} - e^{\beta \hbar \omega /2})(P_{e,e} - P_{g,g})$, where $P_{g,g}$ ($P_{e,e}$) denotes the probability of obtaining $-\hbar \omega /2$ ($\hbar \omega /2$) from both successive measurements performed in the TMA. The exact solution of the model~\cite{QO} allows us to fully determine each trajectory related to the measurements and we obtain $P_{e,e}-P_{g,g} = \sum_{n=0}^{\infty} |a_n|^2 [\cos^2 (\Omega_n T) - \cos^2 (\Omega_{n-1} T)]$, which generally differs from zero leading to a modification of the Bochkov--Kuzovlev identity as described above. We further assume that the initial photonic state is coherent with an average photon number of $\bar{n} = |\alpha|^2$ and define the second measurement to take place at $T=\pi /(2g\sqrt{\bar{n}})$ corresponding to a negative classical injection of a single excitation according to $W_{CL,COH}(t)$ derived earlier. In the limit of large $\bar{n}$, we can approximate $P_{e,e}-P_{g,g} \approx -\frac{\pi}{4\bar{n}} \sum_{n=0}^{\infty} |a_n|^2 \sin \left( \frac{\pi \sqrt{n}}{\sqrt{\bar{n}}} \right)$~\cite{pra44/5913}. Hence, the modification of the Bochkov--Kuzovlev identity scales as $1/\bar{n}$ as we approach the classical-driving limit.

\section{Conclusions}

Our analysis provides a coherent framework for driving and work in quantum systems without resorting to time-dependent external forces. We address the long-standing question of defining work in quantum systems by giving a clear and sound definition for a fully quantized drive. Since the explicit time-dependence of the system Hamiltonian vanishes in our approach, the description of driven dissipative systems~\cite{jsm2009/P02025, pre73/046129} avoids the problem of selecting the appropriate time-dependent basis for defining work~\cite{prl105/030401, pra83/012112, prb84/174507, njp14/123016, pra87/042111}. The theoretical task at hand turns into engineering the drive and its interactions in a manner that allows for the desired complexity in driving protocols. We show that the usual Bochkov-Kuzovlev identity~\cite{spjetp45/125, rmp83/771} is modified by system--drive correlations generated by quantum backaction. To this end, adjusting the measurement protocol in our stochastic analysis possibly provides access to similar modifications of the quantum Jarzynski identity and the integral fluctuation theorem~\cite{rmp81/1665, rmp83/771}. Our work opens a new path in treating driven quantum systems potentially essential for future high-precision experimental studies such as those related to quantum information processing and quantum fluctuation relations. Recent measurements of classical systems~\cite{science296/1832, nature437/231, epl70/593, natphys6/988, prl109/180601, np9/644} and proposals for experiments in classically driven quantum systems~\cite{prl101/070403, prl108/190601, prl110/230601, prl110/230602, njp15/115006, njp15/105028} indicate that such studies are within the grasp of the current technological capabilities.

\acknowledgments

The authors thank M. Campisi, A. Kutvonen, S. Suomela, E.-M. Laine, H. Weimer, T. Ala-Nissil\"a, and J. P. Pekola for useful discussion. This research has been supported by the V\"ais\"al\"a Foundation and the Academy of Finland through its Centres of Excellence Program under Grant No. 251748 (COMP) and Grants No. 138903, No. 135794, and No. 141015. We have received funding from the European Research Council under Starting Independent Researcher Grant No. 278117 (SINGLEOUT). P.S. acknowledges financial support from FIRB-Futuro in Ricerca 2013 under Grant No. RBFR1379UX and FIRB-Futuro in Ricerca 2012 under Grant No. RBFR1236VV HybridNanoDev.

\appendix

\section{Derivation of the injected power and dissipated heating powers} \label{sec:ader}

To construct the various instantaneous powers in accordance with the framework proposed in the main text, we write the energy extracted from the drive as
\begin{widetext}
\begin{equation}
\begin{split}
- \frac{d}{dt}\braket{\hat{H}_D \otimes \hat{I}_S \otimes \hat{I}_E} = -\frac{i}{\hbar} \mathrm{Tr} \{ \hat{\rho} [\hat{H}_{SD} \otimes \hat{I}_E, \hat{H}_D \otimes \hat{I}_S \otimes \hat{I}_E] \} -\frac{i}{\hbar} \mathrm{Tr} \{ [\hat{H}_{DE}, \hat{H}_D \otimes \hat{I}_S \otimes \hat{I}_E] \},
\end{split}
\label{eq:WQ+QD}
\end{equation}
\end{widetext}
where we applied the Ehrenfest theorem~\cite{QM} for time-independent observables $d \braket{\hat{G}}/dt = -(i/\hbar) \mathrm{Tr} \{ \hat{\rho} [\hat{G},\hat{H}] \}$ and $\hat{G}$ denotes an arbitrary observable in the total composite space. In Eq.~(\ref{eq:WQ+QD}), the first term after the equality describes the contribution due to the direct system--drive interaction and the second term gives the corresponding contribution from the direct drive--environment coupling. These terms describe internal and interaction energy changes within the composite and, by energy conservation, we can write a continuity equation in the form of $-d\braket{\hat{H}_D \otimes \hat{I}_S \otimes \hat{I}_E}/dt = dW_Q/dt + dQ_D/dt$. Here we identify $dW_Q/dt$ as the power injected into $S$ and $dQ_D/dt$ as the power directly dissipated from the drive as given by the first and second terms after the equality in Eq.~(\ref{eq:WQ+QD}), respectively. 

By using the internal energy change of the effective inclusive system accounting for both the system $S$ and its drive interaction $SD$, we define the heat directly dissipated from the system, $Q_S$, in accordance with the first law of thermodynamics through
\begin{widetext}
\begin{equation}
\begin{split}
\frac{d}{dt}W_Q - \frac{d}{dt}Q_S &= \frac{d}{dt} \braket{\hat{I}_D \otimes \hat{H}_S \otimes \hat{I}_E + \hat{H}_{SD} \otimes \hat{I}_E} \\ &= -\frac{i}{\hbar} \mathrm{Tr} \{ \hat{\rho} [\hat{I}_D \otimes \hat{H}_S \otimes \hat{I}_E + \hat{H}_{SD} \otimes \hat{I}_E, \hat{H}_D \otimes \hat{I}_S \otimes \hat{I}_E + \hat{I}_D \otimes \hat{H}_{SE} + \hat{H}_{DE}] \}.
\end{split}
\label{eq:WQ-QS}
\end{equation}
\end{widetext}
We identify the total dissipated heat as $Q_{\mathrm{tot}} = Q_S + Q_D$ which, by energy conservation, is equal to the energy extracted from the system-drive composite and, subsequently, has the identity
\begin{widetext}
\begin{equation}
\begin{split}
\frac{d}{dt} Q_{\mathrm{tot}} &= - \frac{d}{dt} \braket{\hat{I}_D \otimes \hat{H}_S \otimes \hat{I}_E + \hat{H}_{SD} \otimes \hat{I}_E + \hat{H}_D \otimes \hat{I}_S \otimes \hat{I}_E} \\ &= -\frac{i}{\hbar} \mathrm{Tr} \{ \hat{\rho} [\hat{I}_D \otimes \hat{H}_{SE} + \hat{H}_{DE},\hat{I}_D \otimes \hat{H}_S \otimes \hat{I}_E + \hat{H}_{SD} \otimes \hat{I}_E + \hat{H}_D \otimes \hat{I}_S \otimes \hat{I}_E] \}.
\end{split}
\label{eq:QTOT}
\end{equation}
\end{widetext}
Executing partial tracing wherever possible in Eqs.~(\ref{eq:WQ+QD}), (\ref{eq:WQ-QS}), and (\ref{eq:QTOT}) yields the definitions for $W_Q$, $Q_S$, and $Q_D$ given in the main text.

\section{Dynamics and injected work under the classical driving assumption} \label{sec:aclas}

The notion of the classical drive is encapsulated by the assumption that the drive dynamics are independent of the other degrees of freedom. As a result, the total evolution factorizes so that $\hat{\rho} = \hat{\rho}_D  \otimes \hat{\rho}_{SE}$ at all times yielding $\mathrm{Tr}_D \{ -(i/\hbar) [\hat{H},\hat{\rho}] \} = -(i/\hbar) [\hat{H}_{CL},\hat{\rho}_{SE}]$ where 
\begin{equation}
\begin{split}
\hat{H}_{CL} = \hat{H}_{CL,S} \otimes \hat{I}_E + \hat{I}_S \otimes \hat{H}_E + \hat{H}_{SE},
\end{split}
\end{equation}
and we write $\hat{H}_{CL,S} = \hat{H}_S + \sum_{\alpha} \hat{A}_{\alpha} \braket{\hat{B}_{\alpha}}_D$ and $\braket{\hat{B}_{\alpha}}_D = \mathrm{Tr}_D \{ \hat{\rho}_D \hat{B}_{\alpha} \}$. We exploit the general decomposition of the interaction Hamiltonian $\hat{H}_{SD} = \sum_{\alpha} \hat{B}_{\alpha} \otimes \hat{A}_{\alpha}$, where $\hat{B}_{\alpha} = \hat{B}_{\alpha}^{\dagger}$ and $\hat{A}_{\alpha} = \hat{A}_{\alpha}^{\dagger}$~\cite{tToOQS}, to write the corresponding effective term $\mathrm{Tr}_D \{ (\hat{\rho}_D \otimes \hat{I}_S \otimes \hat{I}_E) (\hat{H}_{SD} \otimes \hat{I}_E) \}$ and omit $\hat{H}_{DE}$. The term proportional to $\hat{H}_D$ does not appear in $\hat{H}_{CL}$ as it vanishes in the traced commutation. Using the factorization, we can further write $\mathrm{Tr}_D \{ d\hat{\rho}/dt \} = \hat{\rho}_{SE} \mathrm{Tr}_D \{ d\hat{\rho}_D/dt \} + d\hat{\rho}_{SE}/dt$ where $d\hat{\rho}_D/dt = -(i/\hbar) [\hat{H}_D,\hat{\rho}_D]$ by the assumption that the drive dynamics are independent and, hence, $\mathrm{Tr}_D \{ d\hat{\rho}_D/dt \}$ vanishes. Combining the results, we obtain $d\hat{\rho}_{SE}/dt = -(i/\hbar)[\hat{H}_{CL},\hat{\rho}_{SE}]$ determining the system-environment dynamics in classical driving. We neglected $\hat{H}_{DE}$ as the drive-environment interaction is typically not accounted for in the classical picture. Including it would add to $\hat{H}_{CL}$ an effective interaction term $\mathrm{Tr}_{D} \{ (\hat{\rho}_D \otimes \hat{I}_S \otimes \hat{I}_E) \hat{H}_{DE} \} = \sum_{\alpha} \hat{I}_S \otimes \hat{D}_{\alpha} \mathrm{Tr}_D \{ \hat{\rho}_D \hat{C}_{\alpha} \}$ where the final expression uses the decomposition $\hat{H}_{DE} = \sum_{\alpha} \hat{C}_{\alpha} \otimes \hat{I}_S \otimes \hat{D}_{\alpha}$.

Using a power operator method, the classically injected power is derived in Ref.~\cite{prb87/060508} to be $dW_{CL}/dt = d\braket{\hat{H}'_{CL,S}[\lambda(t)]}/dt + dQ_{CL}/dt$ where $\hat{H}'_{CL,S} [\lambda (t)]$ denotes the classically driven system Hamiltonian and $\lambda(t)$ is the time-dependent external control parameter. The first term $\braket{\hat{H}'_{CL,S}[\lambda(t)]}/dt$ describes the internal energy change of the classically driven system and the second term yields the energy exchange with the environment given by the dissipated heat $Q_{CL}$. Identifying $\hat{H}'_{CL,S} [\lambda (t)] = \hat{H}_{CL,S}$, the first term has the explicit form~\cite{prb87/060508}
\begin{widetext}
\begin{equation}
\begin{split}
\frac{d}{dt} \braket{\hat{H}'_{CL,S} [\lambda(t)]} &= -\frac{i}{\hbar} \mathrm{Tr}_{S+E} \{ \hat{\rho}_{SE} [\hat{H}_{CL,S} \otimes \hat{I}_E, \hat{H}_{CL}] \} + \mathrm{Tr}_{S+E} \{ \hat{\rho}_{SE} \partial_t (\hat{H}_{CL,S} \otimes \hat{I}_E) \} \\ &= -\frac{i}{\hbar} \mathrm{Tr}_{S+E} \{ \hat{\rho}_{SE} [\hat{H}_{CL,S} \otimes \hat{I}_E, \hat{H}_{SE}] \} + \mathrm{Tr}_{S+E} \{ \hat{\rho}_{SE} \partial_t (\hat{H}_{CL,S}) \otimes \hat{I}_E \}
\end{split}
\label{eq:dH_CL/dt}
\end{equation}
\end{widetext}
where we applied the Ehrenfest theorem in the effective classical driving picture and $\partial_t$ denotes an explicit time-derivative. Note especially that $\partial_t (\hat{H}_{CL,S}) = \sum_{\alpha} (\partial_t \braket{\hat{B}_{\alpha}}_D) \hat{A}_{\alpha}$. The heat exchange term is given by~\cite{prb87/060508, pra85/032110}
\begin{equation}
\begin{split}
\frac{d}{dt} Q_{CL} = \frac{i}{\hbar} \mathrm{Tr}_{S+E} \{ \hat{\rho}_{SE} [\hat{H}_{CL,S} \otimes \hat{I}_E,\hat{H}_{SE}] \}.
\end{split}
\end{equation}
As described above, the injected power in the full composite framework is $dW_Q/dt = d \braket{\hat{I}_D \otimes \hat{H}_S \otimes \hat{I}_E + \hat{H}_{SD} \otimes \hat{I}_E}/dt + dQ_S/dt$ where the first term after the equality is written as
\begin{widetext}
\begin{equation}
\begin{split}
&\frac{d}{dt} \braket{\hat{I}_D \otimes \hat{H}_S \otimes \hat{I}_E + \hat{H}_{SD} \otimes \hat{I}_E} \\ &= -\frac{i}{\hbar} \mathrm{Tr} \{ \hat{\rho} [\hat{I}_D \otimes \hat{H}_S \otimes \hat{I}_E + \hat{H}_{SD} \otimes \hat{I}_E, \hat{I}_D \otimes \hat{H}_{SE} + \hat{H}_{DE}] \} - \frac{i}{\hbar} \mathrm{Tr} \{ \hat{\rho} [\hat{H}_{SD} \otimes \hat{I}_E, \hat{H}_D \otimes \hat{I}_S \otimes \hat{I}_E] \}.
\end{split}
\end{equation}
\end{widetext}
Applying the classical driving assumption yields
\begin{widetext}
\begin{equation}
\begin{split}
&\frac{d}{dt} \braket{\hat{I}_D \otimes \hat{H}_S \otimes \hat{I}_E + \hat{H}_{SD} \otimes \hat{I}_E} \\ &= -\frac{i}{\hbar} \mathrm{Tr}_{S+E} \{ \hat{\rho}_{SE} [\hat{H}_S \otimes \hat{I}_E + \sum_{\alpha} \braket{\hat{B}_{\alpha}}_D \hat{A}_{\alpha} \otimes \hat{I}_E, \hat{I}_D \otimes \hat{H}_{SE}] \} + \mathrm{Tr}_{S+E} \{ \hat{\rho}_{SE} (\sum_{\alpha} - \frac{i}{\hbar} \mathrm{Tr}_D \{ \hat{\rho}_D [\hat{B}_{\alpha},\hat{H}_D] \} \hat{A}_{\alpha} \otimes \hat{I}_E) \} \\ &= -\frac{i}{\hbar} \mathrm{Tr}_{S+E} \{ \hat{\rho}_{SE} [\hat{H}_{CL,S} \otimes \hat{I}_E, \hat{I}_D \otimes \hat{H}_{SE}] \} + \mathrm{Tr}_{S+E} \{ \hat{\rho}_{SE} \partial_t(\hat{H}_{CL,S}) \otimes \hat{I}_E \},
\end{split}
\end{equation}
\end{widetext}
where we neglect $\hat{H}_{DE}$ and use the factorization of the total evolution after the first equality, and the independence of the drive evolution after the second equality to write $-(i/\hbar) \mathrm{Tr}_D \{ \hat{\rho}_D [\hat{B}_{\alpha},\hat{H}_D] \} = \partial_t \braket{\hat{B}_{\alpha}}_D$. In a similar fashion, the heat dissipated from the system determined by Eqs.~(\ref{eq:WQ+QD}) and (\ref{eq:WQ-QS}) becomes
\begin{widetext}
\begin{equation}
\begin{split}
\frac{d}{dt} Q_S &= -\frac{i}{\hbar} \mathrm{Tr}_{S+E} \{ \hat{\rho}_{SE} [\hat{H}_{SE},\hat{H}_S \otimes \hat{I}_E] \} -\frac{i}{\hbar} \mathrm{Tr} \{ \hat{\rho} [\hat{I}_D \otimes \hat{H}_{SE} + \hat{H}_{DE}, \hat{H}_{SD} \otimes \hat{I}_E] \} \\ &= \frac{i}{\hbar} \mathrm{Tr}_{S+E} \{ \hat{\rho}_{SE} [\hat{H}_{CL,S} \otimes \hat{I}_E,\hat{H}_{SE}] \},
\end{split}
\label{eq:dQ_S/dt_red}
\end{equation}
\end{widetext}
where we neglected $\hat{H}_{DE}$ and used the Born approximation after the second equality. Comparison of Eqs.~(\ref{eq:dH_CL/dt})--(\ref{eq:dQ_S/dt_red}) reveals that $dW_Q/dt = dW_{CL}/dt$ when we identify $\hat{H}'_{CL,S} [\lambda (t)] = \hat{H}_{CL,S}$ and apply the classical driving assumption. Furthermore, the corresponding internal energy change and heating power can be identified as $d \braket{\hat{I}_D \otimes \hat{H}_S \otimes \hat{I}_E + \hat{H}_{SD} \otimes \hat{I}_E}/dt = d \braket{\hat{H}'_{CL,S}[\lambda(t)]}/dt$ and $d Q_S/dt = dQ_{CL}/dt$, respectively.

\section{Interpretation of the modified Bochkov-Kuzovlev identity} \label{sec:aint}

Let us provide a physical interpretation of the result obtained in Eq.~(12) of the main text. We can write $\braket{e^{-\beta W_{\mathrm{excl}}}} = Z_S'(T)/Z_S'(0)$ if we define
\begin{equation}
\begin{split}
Z_S'(t) &= \mathrm{Tr}_{S+D} \{ \hat{U}(t,0) \hat{\rho}_D(0) \otimes \hat{I}_S \hat{U}^{\dagger}(t,0) e^{-\beta \hat{H}_S} \} \\ &= \mathrm{Tr}_S \{ \mathrm{Tr}_D \{ e^{-\beta \hat{H}_S^H(t)} \hat{\rho}_D(0) \} \},
\end{split}
\label{eq:zsre}
\end{equation}
where we used the cyclicity of the full trace and identified $\hat{H}_S^H(t) = \hat{U}^{\dagger}(t,0) \hat{I}_D \otimes \hat{H}_S\hat{U}(t,0)$ as the system Hamiltonian given in the complete Heisenberg picture. Note that unlike $\hat{I}_D \otimes \hat{H}_S$, $\hat{H}_S^H(t)$ operates nontrivially on the drive degrees of freedom and the time-dependence of $Z_S'(t)$ does not indicate any dynamics but characterizes the effective driving protocol. Equation (\ref{eq:zsre}) shows that $Z_S'(t)$ corresponds to averaging the canonical ensemble of the system over the drive degrees of freedom before tracing over the system space, i.e., $Z_S'(t) = \mathrm{Tr}_S \{ \braket{e^{-\beta \hat{H}_S}}_D \}$. Furthermore, by defining $\hat{H}_S^*(t) = -\beta^{-1} \ln \mathrm{Tr}_D \{ e^{-\beta \hat{H}_S^H(t)} \hat{\rho}_D(0) \}$ we can write $Z_S'(t) = \mathrm{Tr}_S \{ e^{-\beta \hat{H}_S^*(t)} \}$ implying that the partition function corresponds to the temporal evolution of the drive-averaged canonical state of the system governed by $\hat{H}_S^*(t)$ which, in this sense, acts as the quantum Hamiltonian of mean force associated with the \textit{bare} system of interest.

The above-mentioned definition of the quantum Hamiltonian of mean force can be compared to the one used in Ref.~\cite{prl102/210401} defined analogously to its classical counterpart~\cite{jsm2004/P09005} if one assumes that the drive subsystem takes on the role of the bath in the bipartite presentation of Ref.~\cite{prl102/210401}. The definitions are inherently different because the previous work assumes that a thermodynamic partition function arising from tracing over the Gibbsian state of the full composite system is well-defined, that is, there is a super-environment in thermal contact with the composite at the preparation stage. In our case, no such contact is assumed and the measurements on the system only give us access to the drive-averaged thermodynamic bare partition function defined in the manner above. Note especially that the definition of the open quantum system partition function $Z_S^{\mathrm{os}} = \mathrm{Tr}_{S+D} \{ e^{-\beta (\hat{I}_D \otimes \hat{H}_S + \hat{H}_{SD} + \hat{H}_D \otimes \hat{I}_S)} \} / \mathrm{Tr}_D \{ e^{-\beta \hat{H}_D \otimes \hat{I}_S} \}$ in Ref.~\cite{prl102/210401} implicitly includes the system--drive interaction energy and, hence, $Z_S'(t)$ is not an extension of $Z_S^{\mathrm{os}}$ to the full composite picture. Starting from $Z_S'(t)$, all relevant drive-averaged thermodynamic quantitities can be defined in the usual manner. For example, the corresponding internal energy for the system $E_S'(t) = -\partial \ln Z_S'(t) / \partial \beta$ obtains the form $E_S'(t) = \mathrm{Tr}_S \{ \mathrm{Tr}_D \{ e^{-\beta \hat{H}_S^H(t)} \hat{H}_S^H(t) \hat{\rho}_D(0) \} \}/Z_S'(t)$
and the averaged free energy is $F_S'(t) = -\beta^{-1} \ln [ \mathrm{Tr}_S \{ e^{-\beta \hat{H}_S^*(t)} \} ]$.

\bibliography{localbib.bib}

\end{document}